\documentclass[proof, usenatbib]{WileyASNA-v1}

\articletype{CONFERENCE PROCEEDING}

\received{19 November 2021}
\revised{19 November 2021}
\accepted{19 November 2021}

\usepackage{hyperref}
\usepackage{xcolor}
\definecolor{citationblue}{HTML}{00657E}

\hypersetup{colorlinks,
            citecolor=citationblue,
            urlcolor=citationblue,
            linkcolor=citationblue}

\pdfoutput=1

\begin{document}

\title{Estimating photoevaporative mass loss of exoplanets with PLATYPOS\protect\thanks{\url{https://github.com/lketzer/platypos}}}

\author[1,2]{Ketzer L.*}

\author[1,2]{Poppenhaeger K.}

\authormark{KETZER \textsc{et al}}

\address[1]{\orgname{Leibniz Institute for Astrophysics (AIP), Potsdam, Germany}}

\address[2]{\orgname{University of Potsdam, Potsdam, Germany}}

\corres{*Ketzer, Laura \email{lketzer@aip.de}}

\presentaddress{\orgname{Leibniz Institute for Astrophysics (AIP)}\\\orgaddress{An der Sternwarte 16, 14482 Potsdam, \country{Germany}}}

\abstract{We develop PLATYPOS (PLAneTarY PhOtoevaporation Simulator), a python code to perform planetary photoevaporative mass-loss calculations for close-in planets with hydrogen-helium envelopes atop Earth-like rocky cores. With physical and model parameters as input, PLATYPOS calculates the atmospheric mass loss and with it the radius evolution of a planet over time, taking into account also the thermal cooling and subsequent radius evolution of the planet. In particular, we implement different stellar activity evolution tracks over time. Our setup allows for a prediction of whether a planet can hold on to a significant fraction of its atmosphere, or fully evaporates, leaving behind only the bare rocky core.
The user supplies information about the star-planet system of interest, which includes planetary and host star parameters, as well as the star's rotational and thus activity evolution. In addition, several details for the evaporative mass-loss rate estimation can be chosen. This includes the effective absorption cross-section for high energy photons, the evaporation efficiency, and the hydrodynamic escape model.}


\maketitle


\section{Introduction}\label{sec1} 

Thanks to the Kepler mission, it is now well established that sub-Neptune-sized planets orbiting their host star with periods less than 100 days are very abundant \citep[e.g.,][]{2010Borucki, 2015Winn}. A second striking discovery regarding these small close-in planets is the uncovered substructure in their radius distribution. Planets tend to group into two distinct populations, the so-called super-Earths and sub-Neptunes, with a significant dearth of planets with intermediate radii around $2\, \mathrm{R}_\oplus$. This had been expected from theoretical studies \citep{2013Owen, Lopez2012} before being observed \citep{Fulton2017, VanEylen2018b, 2021David}.

The gap-like feature, or bimodality in the radius distribution, is predominantly explained by atmospheric erosion of H/He atmospheres caused by the high-energy X-ray and ultraviolet (together: XUV) irradiation from the host star, a process also known as photoevaporation \citep[e.g.,][]{Lopez2012, 2013Owen}; although also alternative scenarios involving core-driven evaporation have been suggested \citep{2018Ginzburg, 2019Gupta}. The planetary properties paired with the external stellar environment, which is determined by host star properties and activity history, impact the strength of the mass loss. If a planet can hold on to a significant fraction of its primordial atmosphere, its radius will be large enough to place the planet above the radius gap. In case of a complete loss of the envelope, only the bare rocky core with a radius below the gap survives. 

The age up to about a Gyr is thought to be most important for the fate of a planet because this is where the most significant mass loss is taking place \citep[e.g.,][]{2013Owen}. Planets still host extended atmospheres because they have not had enough time too cool and contract, and at the same time, they receive the highest XUV flux because young stars can maintain high activity levels. Due to different initial stellar rotation rates, activity levels can vary by about an order-of-magnitude for young stars with similar masses \citep{2011Wright, 2015Tu, 2021Johnstone}. The stellar activity evolution in the first several 100 Myrs thus needs to be taken into account when estimating the planetary mass and radius evolution over time \citep[][Ketzer et al. in prep;]{2021Kubyshkina}.

In this work, we provide a general description of the publicly available code PLATYPOS (PLAneTarY PhOtoevaporation Simulator), which was first applied to the V1298 Tau system \citep{2021Poppenhaeger}. PLATYPOS  is a python code to perform planetary photoevaporative mass-loss calculations for close-in planets with Earth-like rocky cores and H/He envelopes on top. We illustrate some of the capabilities of PLATYPOS in Section\,\ref{sec:example}, using the innermost planet of the V1298 Tau system as an example.

\section{Planetary Evolution Framework}
\label{sec:evo_framework}

PLATYPOS couples a planetary structure model, which includes the planet's thermal evolution, with an atmospheric photoevaporation model to investigate the mass-loss and subsequent radius evolution of a planet over time \citep[as in e.g.,][]{LopezFortney2014, 2017OwenWu}. The code can be used to investigate the mass and radius evolution of individual systems, or be applied to study how atmospheric mass loss shapes a whole population of exoplanets.

PLATYPOS also allows for an easy inclusion of the host star activity evolution in photoevaporation calculations, an important detail which has only recently been incorporated in these types of calculations by, for example, \citet{Kubyshkina2018b, 2021Kubyshkina}.

\begin{figure*}
\centering
\includegraphics[width=0.98\textwidth]{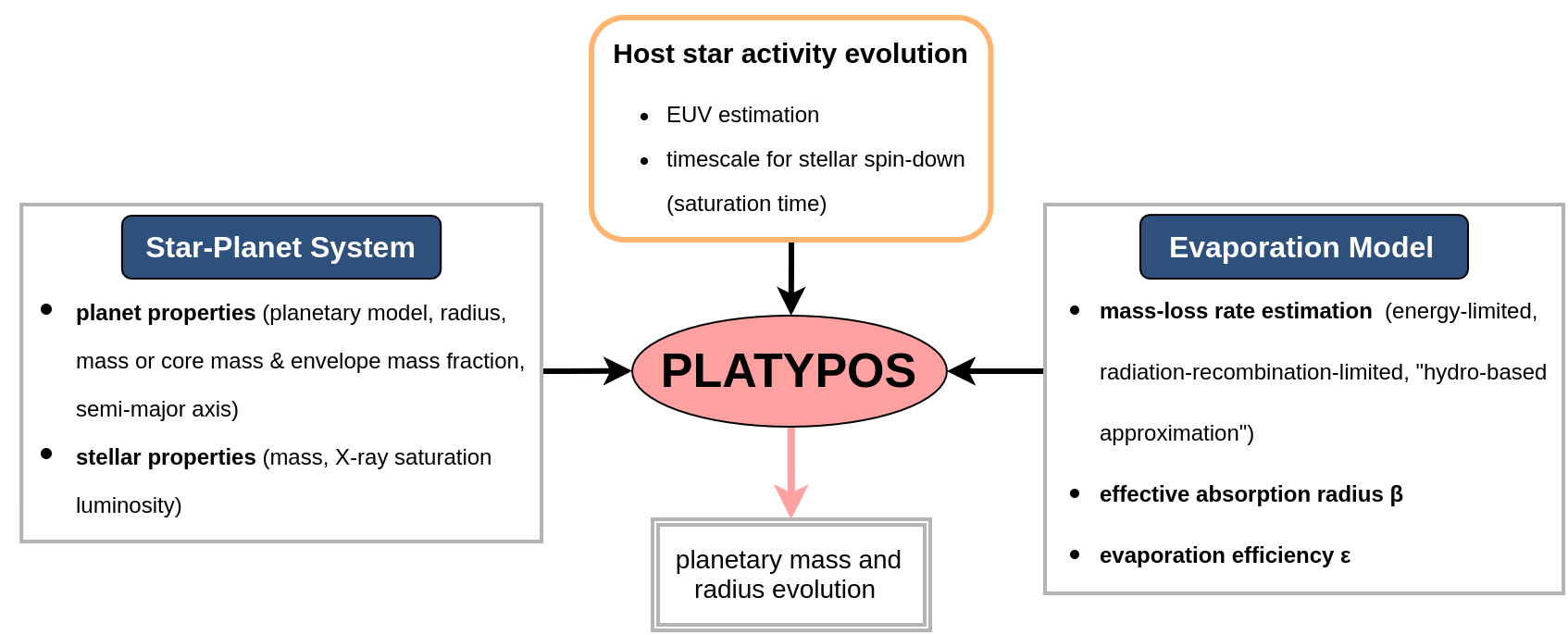}
\caption{Overview of the three main building blocks of PLATYPOS and the individual parameters that need to be measured or estimated to calculate the mass and radius evolution of a planet.}
\label{fig:flowchart}
\end{figure*}

Since photoevaporative mass loss is a complex process, which requires many ingredients, several (simplifying) assumptions need to be made - this includes the planet as well as the host star. The building blocks of PLATYPOS - the planetary models and the mass-loss description, together with additional assumptions about the host star activity evolution - are briefly presented in the following sections. An overview of the ingredients of PLATYPOS is shown in Figure\,\ref{fig:flowchart}.

\subsection{Planetary models}
\label{subsec:planet_models}

To estimate the planetary radius at any given point in time, PLATYPOS has two planetary structure models implemented. The user can choose between the tabulated models by \cite{LopezFortney2014} and the models by \cite{ChenRogers2016}, which are based on the 1-D stellar evoluotion code MESA (hereafter: LoFo and ChRo).
Both models provide mass-radius-age relations for low-mass gaseous sub-Neptune-sized planets, taking into account the cooling and subsequent radius contraction as a planet ages. Fitting formulas for a wide grid of planetary parameters allow PLATYPOS to estimate the planetary radius for a specified core mass, envelope mass fraction, $f_{\mathrm{env}}$, and bolometric incident flux at any given age over the course of the simulation. The user is cautioned that the mass-radius-age relations are only valid for a finite range of planetary parameters, which includes, but is not limited to, planetary age and envelope mass fraction. \cite{LopezFortney2014} showed that the modeled planetary radii can be reasonably backwards extrapolated to ages of $10\,$Myr (see their Figure 2), which is the earliest starting age we recommend to the users of PLATYPOS. This age is a conservative value for the lifetime of a protoplanetary disk \citep[e.g.,][]{2011Williams}: only after its dispersal, the planet is fully exposed to the stellar XUV irradiation.

Regarding the envelope mass fraction, the user can decide whether to extrapolate beyond the lower limit of $0.01\%$, which is reached shortly before planets lose their atmosphere completely, or to continue the calculation keeping the radius constant at the last allowed envelope mass fraction. Our tests across a grid of planets with different parameters showed, however, that in general, if a planet reaches an envelope mass fraction of $f_{\mathrm{env}}=0.01\%$, it cannot hold on to its atmosphere regardless of the radius estimation in the final stages. For the additional details of these models and their applicability, we refer to the original publications.

\subsection{Mass-loss rate calculation}
\label{subsec:evaporation_calcs}

Several regimes of hydrodynamic escape in hydrogen-dominated atmospheres, including energy-limited, radiation-recombination limited, and photon-limited escape, have been identified in theoretical studies \citep[e.g.,][]{2003Lammer, MurrayClay2009, 2012Owen_Jackson, 2016Owen}, and there is recent observational evidence of giant planets supporting these regimes \citep{2021Lampon}. The underlying physics of the escape differs in terms of the production and losses of neutral hydrogen, as well as the processes converting the absorbed stellar radiation into work, which ultimately drives the evaporative outflow. PLATYPOS has different evaporation schemes built in, which allows for an easy comparison against each other. Photoevaporative mass-loss can be estimated using an energy-limited approximation only, including the radiation-recombination limited regime, or via a hydro-based approximation.

In all cases, it is the stellar high-energy X-ray and extreme UV (EUV) radiation, which ionizes and heats the gas in the upper atmosphere. If a significant fraction of the externally supplied energy is converted into work to expand the planetary atmosphere and lift material outside the gravitational well of the planet, mass loss is said to occur in the energy-limited regime. PLATYPOS has the commonly used energy-limited hydrodynamic escape model built-in \cite[see e.g.,][]{OwenJackson2012, Lopez2012}, which assumes that the mass-loss rates are limited by the stellar radiative energy deposition and scale linearly with the high energy incident flux (${\dot{M} \propto \mathrm{F}_{XUV}}$).

In the case of high UV fluxes, the ionization fraction and the temperature of the wind become so high that the material in the upper atmosphere reaches a state of radiation-recombination equilibrium. In this regime, a considerable fraction of the externally supplied (X)UV energy is effectively re-radiated away in the form of Hydrogen Lyman-alpha cooling radiation. This energy sink leads to the mass-loss rates having a shallower dependence on the incoming high energy flux ($\dot{M} \propto \sqrt{\mathrm{F}_{XUV}}$), and can cause the mass loss to be very ineffective for highly irradiated planets \citep[e.g.,][]{MurrayClay2009, Salz2016b}. If chosen by the user, PLATYPOS evaluates both the energy-limited and the radiation/recombination-limited mass-loss rate at each time step of the calculation, and adopts the lesser of the two. This ensures that mass-loss rates are not overpredicted for highly irradiated planets, which is particularly true for young planets orbiting close to their still very active host star. For a more detailed explanation of the radiative/recombinative mass-loss rate calculation implemented in PLATYPOS, see section 2.3 in \citet{2017Lopez}.

The hydro-based approximation is based on the computation of a large grid of hydrodynamic upper atmosphere models \citep{Kubyshkina2018b}. The authors provide analytical expressions for the mass-loss rates as a function of the system parameters based on the grid results. They not only take into account the contribution from high-energy radiation, but also the planetary intrinsic thermal energy and surface gravity. Compared to pure energy-limited mass loss, these mass-loss rates can be orders of magnitudes larger for highly irradiated, low-density planets, and a few factors of 10 lower for more massive planets at larger orbital separations \citep{Kubyshkina2018a}.

Models of escaping atmospheres are extremely complex, and hydrodynamic simulations can predict a wide range of mass-loss rates based on the detailed physics and chemistry included. For this reason, we implement all three of the aforementioned evaporation schemes to give the user the choice to compare them against each other for similarities and differences, and to get a more feasible range of possible mass-loss rates.

\subsection{Effective absorption radius and evaporation efficiency}
\label{subsec:XUV_radius}

Observations as well as hydrodynamic simulations both show that heated and expanded planetary atmospheres can make a planet appear significantly larger when observed in X-ray or EUV compared to optical wavelengths \citep[e.g.,][]{Poppenhaeger2013, Salz2016b}. To obtain reasonable mass-loss rate estimates, the effective XUV absorption radius as well as the evaporation efficiency, need to be estimated for a given planet.

PLATYPOS has two methods implemented for approximating the XUV photosphere of a planet. One is the approximation by \citet{Salz2016b}, which is motivated by results from detailed numerical simulations, while the other is a more theoretical calculation following the arguments presented in \citet{MurrayClay2009}, \citet{ChenRogers2016} and \citet{2017Lopez}. The size of the XUV absorption radius can change significantly for different planet properties, with the gravitational potential playing an important role. In particular, lower-mass planets can host atmospheres, which can be extended up to a few times the optical radius, making them much more susceptible to mass loss. Due to the weak observational constraints on this parameter up to now, we give the user a choice of how to estimate the effective absorption radius for XUV photons. In addition, the user can also choose to set this parameter equal to the optical radius.

In the literature, a wide range of values for the evaporation efficiency parameter, or heating efficiency, have been reported. The values ranges from 0.4 \citep{Lalitha2018} down to 0.01 and even lower for Jupiter-mass planets \citep{Salz2016b}. For planets in the sub-Neptune mass regime, values between 0.1 and 0.3 are commonly used \cite[e.g.,][]{2013Owen, Salz2016b}. PLATYPOS currently requires the user to choose a constant heating efficiency, which is then held constant for the whole duration of the calculation.

\subsection{Host star activity evolution}
\label{sec:activity_decay}

To investigate the atmospheric erosion that planets (might) undergo since their release from the protoplanetary disk, it is important to account for changes in the XUV flux a planet receives over time. Rotational spin-down driven by angular momentum loss via the magnetized stellar wind leads to a decreased stellar activity and with it high-energy XUV over time \citep{1997Guedel, Ribas2005, Booth2017}. This means that young planets receive XUV-irradiation levels that can be several orders of magnitude higher than for the present-day Sun, causing their atmospheres to be hotter, more expanded and susceptible to mass loss.

In addition, the activity level and high-energy emission strongly depends on the rotation rate of the star \citep{2014Reiners}. Stellar rotational evolution models and observations indicate that stars with spectral type F, G, or K start their spin-down earlier than M-dwarfs and at a wide range of ages, with stars born as fast rotators staying active much longer than stars born as slow rotators. This spread in saturation timescales seems to be more pronounced for stars with masses similar to or larger than the Sun \citep[e.g.,][]{2015Tu, 2018Garraffo} and becomes tighter for lower mass stars \citep[e.g.,][]{2021Johnstone, 2020Magaudda}. In general, evolutionary state, activity level, and spectral type all contribute to stars emitting variable amounts of X-ray and UV radiation \citep{2015Chadney}, and should be taken into account when studying the atmospheric mass loss of exoplanets. Exoplanet host stars have been investigated for their current X-ray and extreme-UV emission \citep{Poppenhaeger2010, Monsch2019, Foster2021arXiv}; however, we point out that estimating the past activity history of any given star from present-day measurements is highly non-trivial \citep{2019Kubyshkina}.

\begin{figure}
\includegraphics[trim={0.6cm 0.5cm 1.2cm 1.3cm},clip,width=0.5\textwidth]{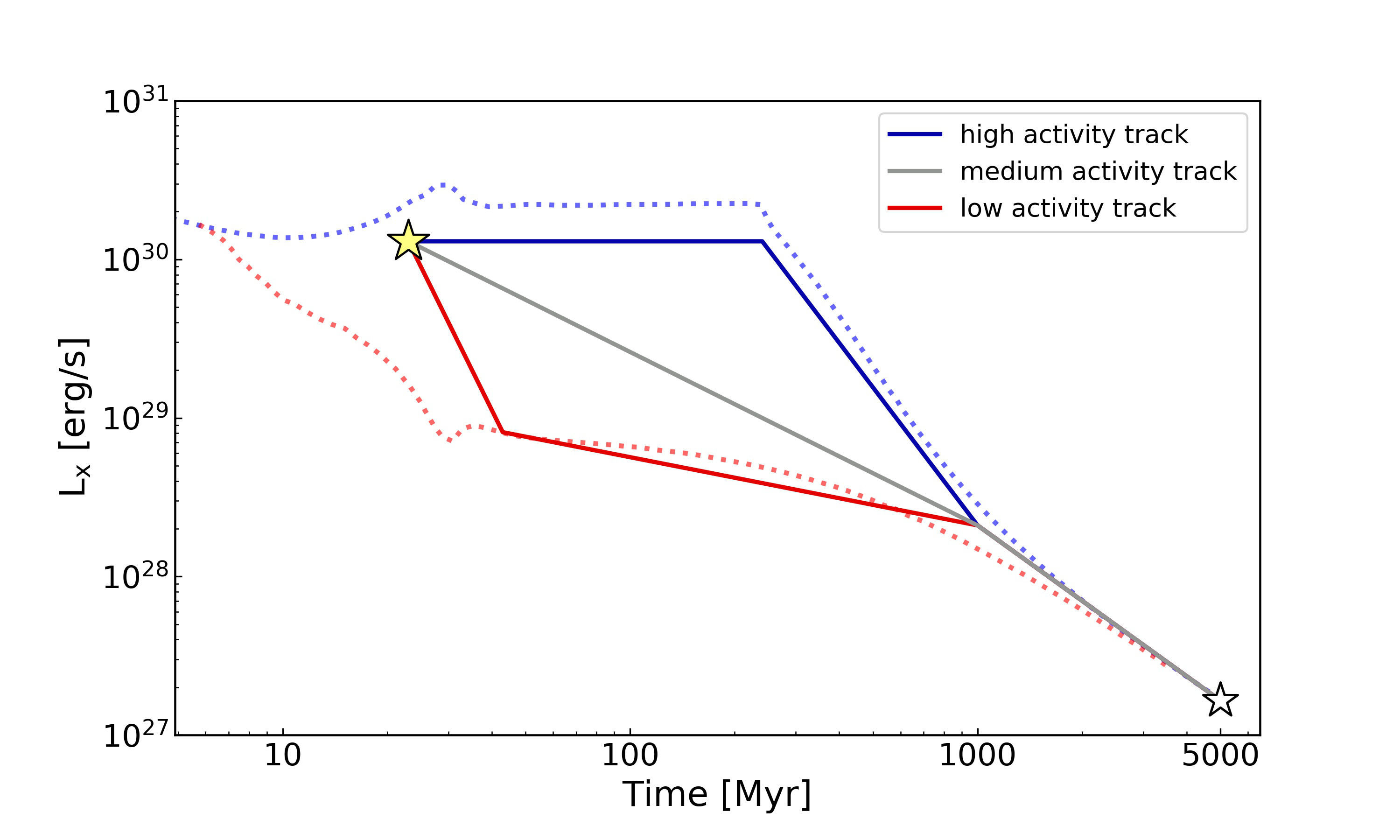}
\caption{A set of example stellar activity tracks for the \mbox{$\sim\,23\,$Myr-}old pre-main sequence star V1298 Tau is shown. The age of the system was determined from isochrone fitting using stellar models that account for magnetic fields \citep{2019David}, and the current X-ray level has been measured with Chandra. It is not well constrained at which age the star will spin down and decrease its activity. We therefore calculated the mass-loss of the planets for a low, intermediate and high stellar activity scenario (red, grey and blue, respectively). For more details see Section 4.2.4 in \citet{2021Poppenhaeger}.}
\label{fig:tracks}
\end{figure}

Currently, the user can choose between the commonly used broken power-law activity evolution with a phase of constant X-ray luminosity, followed by a power-law decay, or a two-piece broken power-law activity decay, which can be used to more realistically simulate X-ray activity tracks with a wide range of saturation or spin-down ages \citep[see][]{2015Tu}. An example of the approximated high, intermediate and low activity evolution tracks used in \citet{2021Poppenhaeger}, together with the detailed model tracks for a fast and slow rotator by \citet{2015Tu}, are shown in Figure\,\ref{fig:tracks}. The code can be easily extended to include any activity track desired by the user.

The important EUV contribution to the high-energy flux can be estimated through the empirical relations between X-ray and EUV surface fluxes by \citet{2015Chadney} or \citet{2021Johnstone}, which have been shown to yield more accurate predictions for active stars and are thought to be valid even on the pre-main sequence. The user can, however, also choose the empirical scaling relation between X-ray and EUV energy bands for late-type stars based on synthetic XUV spectra by \citet{Sanz-Forcada2011}, or the EUV luminosity estimation via Ly$\alpha$ \citep{Linsky2013, Linsky2014}.

\subsection{Details on the integration}

PLATYPOS computes the momentary mass-loss rate for a given planet according to one of the mass-loss formalisms introduced in Section \ref{subsec:evaporation_calcs}, and has estimated the effective absorption radius with one of the methods introduced in Section \ref{subsec:XUV_radius}. It then uses the latest radius, envelope-mass fraction, and stellar XUV flux to calculate the mass-loss rate at the age of the simulation run. Using a fourth-order Runge-Kutta integration method, the mass lost within a given time step is calculated. If the radius change is negligible or too drastic, the time step is adjusted. PLATYPOS then calculates the updated radius with the reduced gaseous envelope based on the planetary model specified by the user (LoFo or ChRo). In the next step, the XUV flux is updated based on the specified stellar evolution track and EUV estimation method, and this cyclic procedure continues until the planetary radius has reached the core radius and no atmosphere remains, or the final age of the simulation is reached. The temporal mass and radius evolution caused by planetary cooling and atmospheric photoevaporation for the specified stellar activity track is then saved. A big advantage of PLATYPOS is that it enables the user to easily change and compare various model assumptions, and to investigate their impacts on the strength of the mass loss and the fate of the planet of interest.

\subsection{Code limitations}

PLATYPOS does not make use of complex radiative-hydrodynamical simulations, but instead brings together parametrized models for the planetary structure, the atmospheric escape as well as the stellar activity evolution for a quick and easy-to-use estimation of planetary photoevaporative mass loss.
All in all, the tool is relatively simple and does not seek to include all potentially relevant physical aspects of exoplanet evaporation. Examples for effects not considered are interactions of the stellar wind with the planetary outflow, any magnetic shielding effects due to a planet's magnetic field, or any hydrodynamic effects. In addition, the evaporation efficiency is taken to be constant for the whole duration of the calculation. This has been shown to be an oversimplification since the parameter depends on planetary mass, radius, and the amount of ionizing flux, quantities, which can vary by orders of magnitude over the lifetime of the star-planet system \citep[see e.g.,][]{2013Owen}. Various simulations, however, indicate that values around 0.1-0.2 are reasonable for low-mass planets in the super-Earth and sub-Neptune regime \citep{Lopez2013, 2013Owen, Salz2016b}.

Despite these limitations, PLATYPOS makes it feasible to visualize how even the inclusion of a few physical parameters can significantly alter the predicted future mass and radius evolution of a planet. This includes the stellar activity evolution, and with it a star's X-ray saturation luminosity, spin-down behavior, and EUV emission, but also planetary parameters like core mass or initial envelope mass fraction. A third important component is the evaporation model, in particular, the mass-loss rate estimation and the effective XUV absorption radius. In the future, more observations are needed to put tighter constraints on theoretical models and their underlying assumptions.

\section{\mbox{Planet V1298\,Tau\,c} as an example}\label{sec5}
\label{sec:example}

We use the innermost planet of the V1298 Tau system to illustrate how PLATYPOS can be used to explore how some of the underlying assumptions regarding the planet and the evaporation-model details impact the mass-loss predictions. In \citet{2021Poppenhaeger}, we estimated the fate of the four V1298 Tau planets using the LoFo models, energy-limited mass loss, and the EUV estimation by \citet{Sanz-Forcada2011}, and showed that the stellar activity track also plays a major role in determining whether a close-in planet can hold on to some atmosphere or will evaporate completely. More details on the role of the host star in photoevaporation population studies will be discussed in Ketzer et al. (in prep.).

\begin{figure}
\centering
\includegraphics[trim={0.4cm 1.2cm 0 0.9cm},clip,width=0.5\textwidth]{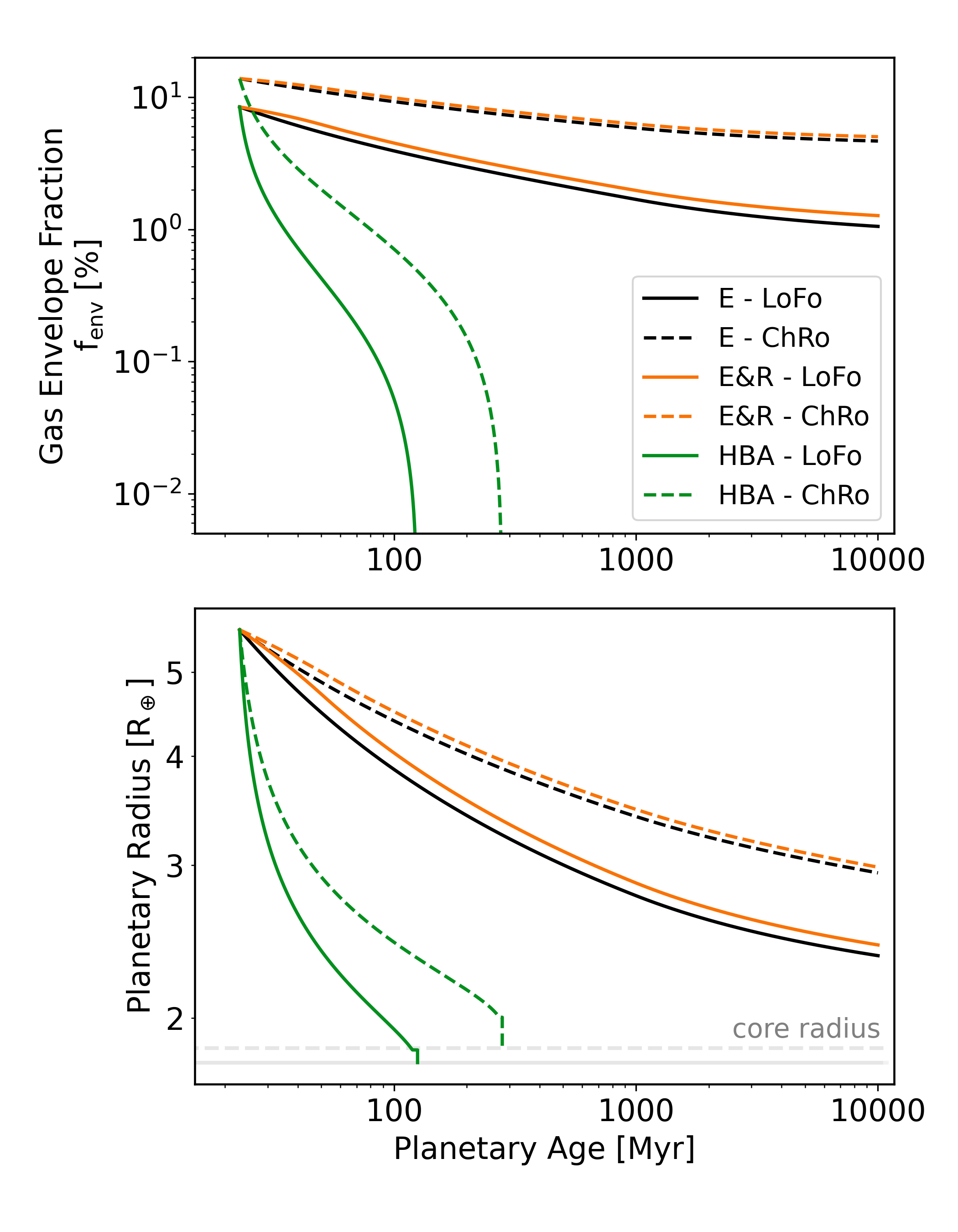}
\caption{Results of our example calculation for the planet V1298 Tau c assuming a 10 Earth-mass core and the medium stellar activity track (gray) from Figure\,\ref{fig:tracks}. The evolved planets all match the observed radius at 23 Myr, the starting age of the simulation. The top panel shows the evolution of the envelope mass fraction, the bottom panel the corresponding radius evolution. We show how the three different evaporation models, as well as the choice of the planetary model, impact the fate of the planet. The energy-limited approximation (E) is shown in black, the evaporation including a radiation/recombination-limited regime (E\&R) in orange, and the hydro-based approximation (HBA) in green. The solid lines are for the LoFo model, while the dashed lines represent the ChRo model. The grey lines indicate the radius for a 10 Earth-mass core as predicted by the LoFo and ChRo models. The sharp drop in radius for the green tracks arises because the planets, after having reached an envelope mass fraction of 0.01\%, evaporate completely within the next, 0.01 Myr-short, timestep.}
\label{fig:V1298Tau}
\end{figure}

We highlight here that the exact details of the planetary structure model affect how quickly a planet cools and contracts. This directly influences the amount of atmosphere needed to match the observed planetary radius at the current age of the system. For a 10 Earth-mass core, using the LoFo model, the innermost planet of V1298 Tau requires an envelope mass fraction of about 8\%, while for the ChRo models this value is 14\% at an age of 23 Myr. In general, a more massive atmosphere is more compact and thus less susceptible to mass loss. As a direct consequence, the ChRo planet will end up with a more massive atmosphere compared to the LoFo one for the same evaporation model and activity track (see Figure\,\ref{fig:V1298Tau}). Nonetheless, the prediction that the planet can hold on to enough atmosphere to stay above the gap is the same for both planet models under the assumption of a medium activity track and the energy-limited evaporation model. The beta and EUV estimations were calculated as described in \citet{2021Poppenhaeger}.

The second observation regards the choice of the evaporation model. For the most irradiated planet in the system, the difference between the energy-limited mass loss and the inclusion of a radiation/recombination-limited regime does not change the final results significantly. We only show the calculation for the medium activity track in Figure\,\ref{fig:V1298Tau}, but the results look qualitatively similar for the low and high activity track. However, the difference in the amount of mass lost between the two evaporation models becomes larger going from a low to high activity track. This is due to the fact that for a prolonged phase of high irradiation levels, the planet is able to cool more efficiently through radiative cooling and thus lose less atmosphere compared to energy-limited mass loss only. A striking difference, however, can be seen when comparing energy-limited (or radiation/recombination-limited) mass loss to the hydro-based calculation. The predicted initial hydro mass-loss rates are more than an order of magnitude higher, which means the planet will lose it's atmosphere within less than 300 Myrs (see Figure\,\ref{fig:V1298Tau}).

\begin{figure}
\centering
\includegraphics[trim={0.4cm 1.2cm 0 0.9cm},clip,width=0.5\textwidth]{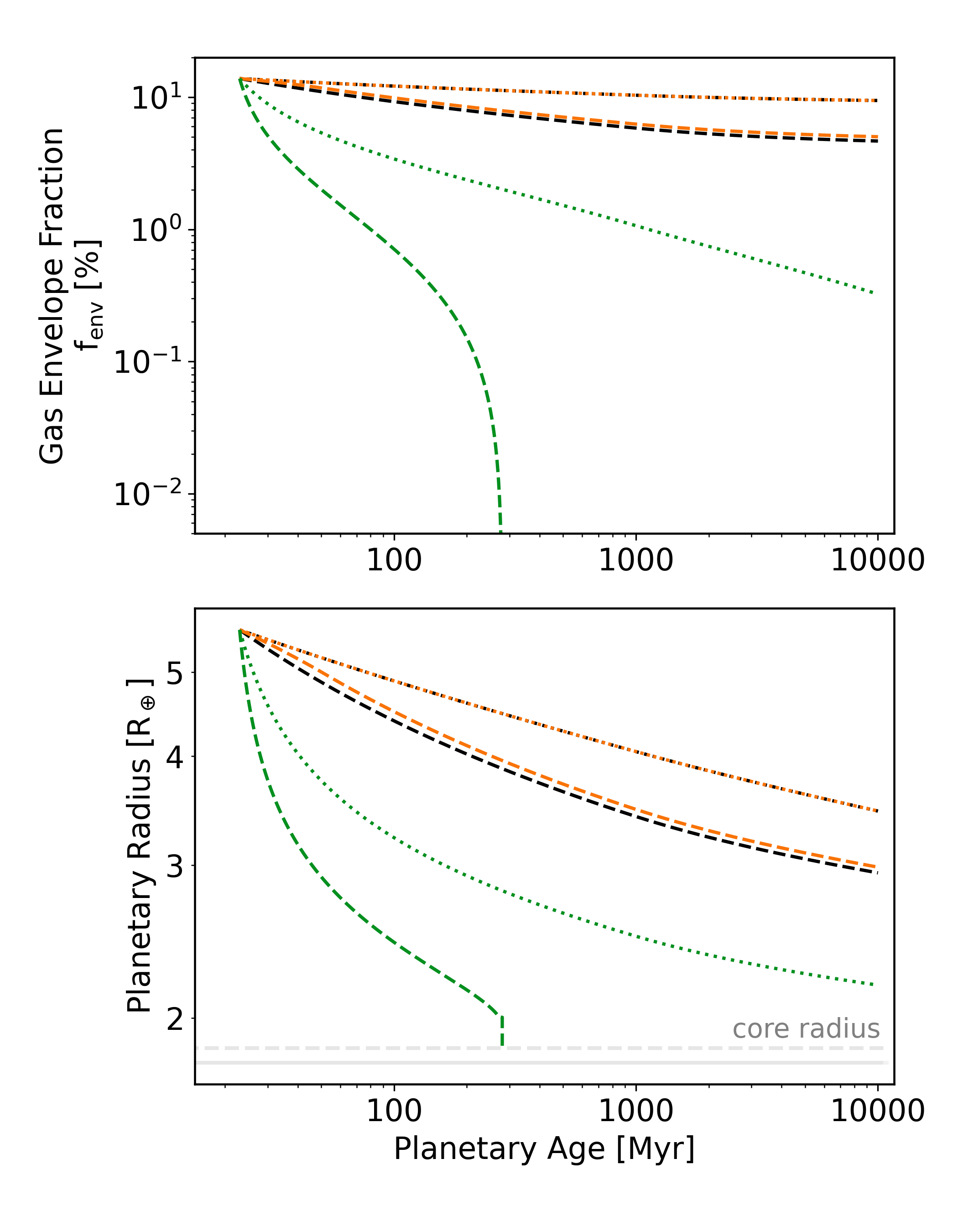}
\caption{Results for the ChRo planet model and all three mass-loss rate estimation methods. The dashed lines are the same as in Figure\,\ref{fig:V1298Tau}, making use of the EUV estimation from \citet{Sanz-Forcada2011}, while the dotted lines represent the evolution for envelope mass fraction and radius for the EUV estimation by \citet{2021Johnstone}, which predicts EUV fluxes that are about a factor 4 lower.}
\label{fig:EUV_impact}
\end{figure}

The EUV estimation can also have a large impact on the final results. In general, the EUV estimation based on surface fluxes by \citet{2021Johnstone} predicts fluxes a few factors lower than the EUV estimation method by \citet{Sanz-Forcada2011}. For V1298 Tau c, this means that the mass loss, in particular in the early stages, is less detrimental and more atmosphere can survive. For the LoFo planet, the final fate of the planet is unchanged for all mass-loss calculations, but for the ChRo planet with the more massive initial atmosphere, the lower EUV irradiation from the "Johnstone"-estimation method leads to the planet surviving with about 1\% of atmosphere even for the hydro-based mass-loss calculation (see Figure\,\ref{fig:EUV_impact}). This result stresses the importance of having a good handle on the X-ray and EUV luminosity of a star in the first Gyr or so, when the strongest mass-loss is occurring.

Ultimately, not only the details of the mass-loss rate estimation, like evaporation model, effective absorption cross section or heating efficiency, can make a large difference in the predicted fate of a planetary atmosphere, but also the stellar evolution track, as well as the amount of X-ray and EUV emission from the host star, can change the calculation results significantly. Detailed predictions of the influence of individual parameters in the model across a wide range of stellar and planetary parameters can be complicated due to the large number of partly intertwined model parameters. More observations of escaping atmospheres are needed to put tighter constraints on mass-loss models and to decide if the mass-loss rates are indeed as high as predicted by the hydro-approximation. In addition, more detailed simulations of the interaction with the stellar wind or planetary magnetic shielding can help to determine the true strength of the mass-loss in the first few 100 Myrs \citep[see, e.g][]{2021Carolan}. We stress that among all evaporation details, the host star and its level of X-ray and EUV emission in the saturated phase, as well as the timescale for the activity decay should not be neglected in photoevaporation studies.

\section{Summary}
We present PLATYPOS, a publicly available Python code to assess the atmospheric mass loss due to XUV irradiation of planets in the super-Earth and sub-Neptune regime. The code makes it easy to estimate the future mass and radius evolution of a young planet, and to explore the impact of the evaporation model details or the stellar activity evolution on the fate of a planet.

\section{Acknowledgments}

We would like to thank the referee for constructive feedback. The scientific results reported in this article are based in part on observations made by the Chandra X-ray Observatory. This work was supported by the German \textit{Leibniz-Gemeinschaft}, project number P67-2018.

\bibliography{Wiley-ASNA}

\end{document}